\documentclass[11pt]{article}
\usepackage{amssymb,latexsym,amsmath}
\usepackage[dvips]{graphicx}
\newtheorem{theo}{Theorem}

\newtheorem{prop}[theo]{Proposition}
\def\ii{\mathrm{i}} 
\def\ee{\mathrm{e}} 
\newcommand{\myatop}[2]{\genfrac{}{}{0pt}{}{#1}{#2}}
\begin{document}
\begin{center}
{\Large \bf
The oscillator model for the Lie superalgebra $\mathfrak{sh}(2|2)$\\[2mm]
and Charlier polynomials}\\[5mm]
{\bf E.I.\ Jafarov\footnote{Permanent address: 
Institute of Physics, Azerbaijan National Academy of Sciences, Javid av.\ 33, AZ-1143 Baku, Azerbaijan}
and J.\ Van der Jeugt} \\[1mm]
Department of Applied Mathematics and Computer Science,
Ghent University,\\
Krijgslaan 281-S9, B-9000 Gent, Belgium\\[1mm]
E-mail: ejafarov@physics.ab.az, Joris.VanderJeugt@UGent.be
\end{center}

\vskip 10mm
\noindent
Short title: $\mathfrak{sh}(2|2)$ oscillator model and Charlier polynomials



\begin{abstract}
We investigate an algebraic model for the quantum oscillator based upon the Lie superalgebra $\mathfrak{sh}(2|2)$,
known as the Heisenberg-Weyl superalgebra or ``the algebra of supersymmetric quantum mechanics'', 
and its Fock representation.
The model offers some freedom in the choice of a position and a momentum operator, leading to a free
model parameter $\gamma$.
Using the technique of Jacobi matrices, 
we determine the spectrum of the position operator, and show that its wavefunctions are related
to Charlier polynomials $C_n$ with parameter $\gamma^2$. 
Some properties of these wavefunctions are discussed, as well as some other 
properties of the current oscillator model.
\end{abstract}

\section{Introduction}

The relation between (one-dimensional) quantum oscillator models and orthogonal polynomials (of a single variable) is very deep.
For the canonical quantum oscillator, it can be found in any textbook that the position wavefunction is given in terms
of Hermite polynomials.
Among the first alternative oscillator model is the paraboson oscillator~\cite{Wigner,Green53}, where the position 
wavefunctions are given in terms of Laguerre polynomials~\cite{Mukunda,JLV2008}.
Depending on the oscillator model under consideration, the spectrum of the position operator can be continuous or 
discrete (with finite or infinite support), and so the wavefunction can correspond to a continuous or discrete 
orthogonal polynomial.
Another example of an oscillator model with a continuous position spectrum is the $\mathfrak{su}(1,1)$ oscillator studied
by Klimyk~\cite{Klimyk2006}, where the wavefunctions are given in terms of Meixner-Pollaczek polynomials, 
or its deformation the $su(1,1)_\gamma$ oscillator~\cite{JSV2012} with wavefunctions given in terms of
continuous dual Hahn polynomials. 
Note that for all of the examples mentioned there is an algebra underlying the model, namely the Heisenberg-Weyl algebra,
the paraboson algebra $\mathfrak{osp}(1|2)$, the Lie algebra $\mathfrak{su}(1,1)$ and its deformation $\mathfrak{su}(1,1)_\gamma$, respectively.

In recent years there was also more interest in so-called finite or discrete oscillator models, inspired by
applications in quantum optics or signal analysis~\cite{Atak2005}. 
The seminal paper in this area is~\cite{Atak2001}, where an oscillator model based on the Lie algebra $\mathfrak{su}(2)$ was
given, with position wavefunctions expressed in terms of symmetric Krawtchouk polynomials. 
This model was extended by an extra parameter in~\cite{JSV2011,JSV2011b}, with an underlying
deformed $\mathfrak{su}(2)$ algebra, and position wavefunctions given by means of Hahn polynomials. 

Lately, also {\em Lie superalgebras} came into the picture as algebras underlying oscillator models.
In~\cite{JV2012}, the Lie superalgebra $\mathfrak{sl}(2|1)$ and its finite-dimensional representations were used to give an oscillator model with general Krawtchouk polynomials as wavefunctions (so with a discrete and finite position spectrum).
Following this, infinite-dimensional discrete series representations of $\mathfrak{sl}(2|1)$ were constructed
in~\cite{Jafarov2012}, and this gave rise to an oscillator model with Meixner polynomials as wavefunctions 
(with a discrete and infinite position spectrum).
The $\mathfrak{sl}(2|1)$ oscillator model with discrete series representations is of particular physical relevance, since it contains -- 
as special cases (i.e.\ for special values of the parameters) -- the paraboson oscillator and thus also the canonical quantum oscillator. 
In this sense, this Lie superalgebraic oscillator model forms a natural extension of the known oscillators.

After completing~\cite{Jafarov2012}, we realized that an even simpler Lie superalgebraic oscillator model could be proposed.
This is the subject of the current paper.
The algebraic context is provided by the Heisenberg-Weyl superalgebra $\mathfrak{sh}(2|2)$, an algebra generated by one boson and 
one fermion pair. This algebra is usually cited in the framework of supersymmetric quantum mechanics.
We manage to construct a simple oscillator model based on this Lie superalgebra, for which the position wavefunctions correspond
to {\em Charlier polynomials}. 
Although the direct physical relevance of this model may be less obvious than the $\mathfrak{sl}(2|1)$ model,
we believe it is worth publishing because of its mathematical beauty and simplicity.
Quantum oscillator models are the fundamental tools in many branches of physics, 
so also the $\mathfrak{sl}(2|1)$ ``toy model'' presented here deserves attention.

In the following section, we shall introduce the superalgebra $\mathfrak{sh}(2|2)$, and more precisely a subalgebra ${\cal S}$ of its
universal enveloping algebra. All (anti)commutation relations of this algebra are given, together with the action of
its basis elements on the Fock space representation $V$ of $\mathfrak{sh}(2|2)$.
With the purpose of describing an oscillator model, we study the spectrum and the formal eigenvectors of a general
self-adjoint odd element $\hat q$ of ${\cal S}$ in section~3. This is done in a broader context, since there has been
some interest in the general construction of so-called superalgebra eigenstates~\cite{Alvarez-Moraga}.
The spectral analysis of this operator $\hat q$ leads to a Jacobi matrix, and yields the infinite discrete spectrum 
$\mathbb{S}= \{\ldots,-\sqrt{3},-\sqrt{2}, -1,0,1,\sqrt{2},\sqrt{3}, \ldots\}$ for $\hat q$, with eigenvectors
given in terms of Charlier polynomials. 
In section~4 we turn to the $\mathfrak{sh}(2|2)$ oscillator model: this is an algebraic model for which the Hamiltonian, the
position and the momentum operators are elements of $\mathfrak{sh}(2|2)$ (or ${\cal S}$). The position operator is just the operator $\hat q$
studied in section~3. So we do indeed come up with an oscillator model for which the position wavefunctions are
given by Charlier polynomials. In that section, we also present some properties of plots of these wavefunctions,
and discuss a limit relation to another oscillator model. 
Since the momentum eigenvectors are also easy to give for the $\mathfrak{sh}(2|2)$ model, it is a simple task to 
construct the corresponding $\mathfrak{sh}(2|2)$ Fourier transform kernel explicitly.
The section ends with the computation of some other physical quantities of the model, such as uncertainty relations.
In the final section, we give a differential operator realization of the superalgebra ${\cal S}$, and 
end with some conclusions.

It should be noted that Charlier polynomials have been related before to the Weyl algebra (and thus -- in a certain sense -- 
to the ordinary oscillator), e.g.\ in~\cite{Floreanini1993,Vinet2011}. 
In particular, they appear as overlap coefficients between eigenvectors of a number operator and a ``shifted'' number operator~\cite{Vinet2011}. 
In the current paper their appearance is very different, and they are, to our knowledge, for the first time related to 
position wavefunctions of a (non-canonical) oscillator.

Finally, in this introduction we have spent some emphasis on the relation between algebraic oscillator models and
related orthogonal polynomials appearing as wavefunctions. We should of course mention that also in the
context of $q$-deformations of oscillator models and $q$-orthogonal polynomials, a large amount of work
has been performed, see~\cite{Klimyk2005} and references therein.

\section{The Lie superalgebra ${\cal S}$ of the supersymmetric oscillator and its Fock representation}

The supersymmetric version of the one-dimensional harmonic oscillator has been studied by many authors, mainly in
the context of supersymmetric quantum mechanics~\cite{Witten1981, Witten1982, DeCrombruggheRittenberg, DHokerVinet, Beckers1987}.
In~\cite{Beckers1987,Beckers1988}, the emphasis is on determining superalgebras of symmetries and supersymmetries, and relations
to known Lie superalgebras have been studied. 
Let us reconsider here the case of a one-dimensional supersymmetric oscillator, the simplest example of $N=2$
supersymmetric quantum mechanics. 
Such an oscillator is realized in terms of a bosonic creation and annihilation operator $b^\pm$ with commutation relation
\begin{equation}
[b^-,b^+]=1,
\label{bb}
\end{equation}
and a fermionic creation and annihilation operator $a^\pm$ with anticommutation relations
\begin{equation}
\{a^-,a^-\}=\{a^+,a^+\}=0, \qquad \{a^-,a^+\}=1.
\label{aa}
\end{equation}
Furthermore, the two sets of operators commute with each other:
\begin{equation}
[b^\xi, a^\eta]=0, \qquad \xi,\eta \in \{+,-\}.
\label{ba}
\end{equation}
The Lie superalgebra generated by the even elements $1,b^+,b^-$ and the odd elements $a^+, a^-$ is known as the
Heisenberg-Weyl superalgebra $\mathfrak{sh}(2|2)$.
The algebra of symmetries or the invariance superalgebra of the supersymmetric oscillator is usually identified as
a subalgebra of the enveloping algebra ${\cal U}(\mathfrak{sh}(2|2))$. 
Let us consider here the Lie superalgebra ${\cal S} \subset {\cal U}(\mathfrak{sh}(2|2))$ with four odd basis elements
\begin{equation}
F^+=a^+,\quad F^-=a^-,\quad Q^+=b^+a^-,\quad Q^-=b^-a^+, \label{evengen}
\end{equation}
and even basis elements
\begin{equation}
E^+=b^+,\quad E^-=b^-,\quad H=b^+b^-+a^+a^- \hbox{ and } 1.  \label{oddgen}
\end{equation}
As a supersymmetric quantum system, $H$ is the Hamiltonian and $Q^\pm$ are the supercharges commuting with $H$ and satisfying
$\{Q^+,Q^-\}=H$.

It is important to note that the Lie superalgebra ${\cal S}$ closes under the Lie superalgebra bracket (without the addition
of extra quadratic operators to ~\eqref{evengen} and~\eqref{oddgen}).
Indeed, the complete set of (anti)commutation relations for the odd elements of ${\cal S}$ is given by:
\begin{align}
& \{F^\pm,F^\pm\}= \{Q^\pm,Q^\pm\}=0, \quad \{F^+,F^-\}=1,\quad \{Q^+,Q^-\}=H, \nonumber\\
& \{F^\pm,Q^\mp\}=0, \quad \{F^+,Q^+\}=E^+,\quad \{F^-,Q^-\}=E^- .
\label{oddodd}
\end{align}
The even-even commutation relations read 
\begin{equation}
[E^-,E^+]=1,\quad [H,E^\pm]=\pm E^\pm,
\end{equation}
where of course 1 is a central element commuting with all elements.
Finally, the even-odd commutation relations are
\begin{align}
& [E^\pm,F^\pm]=[E^\pm,F^\mp]=0, \quad [E^\pm,Q^\pm]=0,\quad [E^\pm,Q^\mp]= \mp F^\pm,\nonumber\\
& [H,F^\pm]=\pm F^\pm,\quad [H,Q^\pm]=0.
\end{align}

The common Fock representation $V$ of $\mathfrak{sh}(2|2)$, generated by a vacuum vector $|0\rangle$ and by the relations
\[
b^-|0\rangle = a^-|0\rangle =0
\]
carries also an irreducible representation of the Lie superalgebra ${\cal S}$. An orthonormal basis of $V$ is given by the 
vectors
\[
\frac{(b^+)^m}{\sqrt{m!}} (a^+)^j |0\rangle, \qquad m\in\{0,1,2,\ldots\}, \ j \in\{0,1\}.
\]
It will be convenient to write these vectors as $|n\rangle$, with $n=0,1,2,\ldots$, so $V$ can be identified with 
the Hilbert space $\ell^2({\mathbb Z}_+)$:
\begin{equation}
|2m\rangle = \frac{(b^+)^m}{\sqrt{m!}} |0\rangle, \qquad |2m+1\rangle = \frac{(b^+)^m}{\sqrt{m!}} a^+ |0\rangle.
\end{equation}
Now it is straightforward to determine the action of the odd elements of ${\cal S}$ on this basis:
\begin{align}
& F^+ |n\rangle = \frac{1}{2}(1+(-1)^n) |n+1\rangle, \qquad F^-|n\rangle = \frac{1}{2}(1-(-1)^n) |n-1\rangle, \nonumber\\
& Q^+ |n\rangle = \frac{1}{2}(1-(-1)^n)\sqrt{\frac{n+1}{2}} |n+1\rangle,\qquad
 Q^- |n\rangle = \frac{1}{2}(1+(-1)^n)\sqrt{\frac{n}{2}} |n-1\rangle.
 \label{act1}
\end{align}
Similarly, one finds for the action of the even elements:
\begin{align}
& E^+ |n\rangle = \frac{1}{2}\left( (1+(-1)^n)\sqrt{\frac{n+2}{2}} + (1-(-1)^n)\sqrt{\frac{n+1}{2}}\right) |n+2\rangle,\nonumber\\
& E^- |n\rangle = \frac{1}{2}\left( (1+(-1)^n)\sqrt{\frac{n}{2}} + (1-(-1)^n)\sqrt{\frac{n-1}{2}}\right) |n-2\rangle,\nonumber\\
& H |n\rangle = \left( \frac{n}{2}+\frac{1-(-1)^n}{4}\right) |n\rangle, \qquad 1 |n\rangle = |n\rangle.
\label{act2}
\end{align}
This Fock representation is unitary with respect to the $\star$-structure
\begin{equation}
(F^\pm)^\dagger= F^\mp, \ (Q^\pm)^\dagger=Q^\mp, \ (E^\pm)^\dagger=E^\mp,\ H^\dagger=H,
\label{dagger}
\end{equation}
which follows from the common conjugacy relations for the boson/fermion creation and annihilation operators $b^\pm$ and $a^\pm$.

To conclude this section, note that in the following we will not be dealing with a supersymmetric oscillator directly, but with
an algebraic oscillator model related to the Lie superalgebra ${\cal S}$.
This is also different from the oscillator based on a deformed Heisenberg algebra~\cite{P1,P2}, which coincides (from the
algebraic point of view) with the paraboson oscillator described by the Lie superalgebra $\mathfrak{osp}(1|2)$.

\section{Eigenstates of a self-adjoint odd element of ${\cal S}$}

The algebra eigenstates associated to a real Lie algebra have been introduced by Brif~\cite{Brif1997}; they can be considered
as generalizations of coherent or squeezed states.
In~\cite{Alvarez-Moraga}, some superalgebra eigenstates of the Heisenberg-Weyl superalgebra $\mathfrak{sh}(2|2)$ have been computed,
and properties of these states were investigated.
In the current section, we shall determine the eigenstates (and eigenvalues) of a general self-adjoint odd element of 
the real Lie superalgebra ${\cal S}$.
Given the conjugations~\eqref{dagger}, the most general self-adjoint odd element, up to an overall factor, is
\begin{equation}
\hat q = \gamma F^+ + Q^+ + \gamma F^- + Q^-,
\label{hatq}
\end{equation}
where $\gamma$ is some real number.
In the ordered basis $\{ |n\rangle, \; n=0,1,2,\ldots\}$ of the Fock space $V$, the operator $\hat q$ is
represented by an infinite symmetric tridiagonal matrix:
\begin{equation}
\hat q=\left(
\begin{array}{cccccccc}
0 & \gamma & & & & & &  \\
\gamma & 0 & \sqrt{1} & & & & & \\
 & \sqrt{1} & 0 & \gamma & & & & \\
 & & \gamma & 0 & \sqrt{2} & & & \\
 & & & \sqrt{2} & 0 & \gamma & & \\
 & & & & \gamma & 0 & \sqrt{3} & \\
 & & & & & \sqrt{3} & 0 & \ddots  \\ 
 & & & & & & \ddots & \ddots \\
\end{array}
\right) .
\label{Mq}
\end{equation}
The form of this matrix follows from the actions~\eqref{act1}-\eqref{act2}.
For $\gamma>0$, such a matrix is a Jacobi matrix, and its spectral theory is related to 
orthogonal polynomials~\cite{Berezanskii,Koelink1998,Koelink2004}. 
It will soon be clear that only $\gamma^2$ plays a role in the eigenvalues and eigenvectors of $\hat q$,
so let us assume for the moment that $\gamma>0$ (the case $\gamma=0$ is trivial, since the matrix~\eqref{Mq} decomposes 
into irreducible $2\times 2$ blocks).
The procedure to deal with such matrices is described in~\cite[\S 2]{Koelink1998}.
One should construct polynomials $p_n(x)$ of degree $n$
in $x$, with $p_{-1}(x)=0$, 
$p_0(x)=1$, and
\begin{align}
&x p_{2n}(x) = \sqrt{n} p_{2n-1}(x) + \gamma p_{2n+1}(x), \nonumber \\
&x p_{2n+1}(x) = \gamma p_{2n}(x) + \sqrt{n+1} p_{2n+2}(x), \qquad (n=0,1,2,\ldots).
\label{def-px}
\end{align}
Such polynomials are orthogonal for some positive weight function $w(x)$, and the spectrum of $\hat q$ 
is the support of this weight function.
There is a mathematical condition related to this: the so-called (Hamburger) moment problem
for the Jacobi matrix should be determinate~\cite{Berezanskii,Koelink2004}. 
For a general tridiagonal Jacobi matrix, a sufficient condition~\cite{Berezanskii,Koelink2004} for this is that
the (infinite) sum of the matrix elements above the diagonal should be infinite,
which is clearly satisfied for~\eqref{Mq}.
So the spectrum of $\hat q$ is just the support of the weight function $w(x)$.
Furthermore, for a real value $x$ belonging to this support, the corresponding formal eigenvector of $\hat q$ is given by
\begin{equation}
v(x) = \sum_{n=0}^\infty p_n(x) \,|n\rangle.
\label{vx}
\end{equation}

The recurrence relations~\eqref{def-px} lead to a solution of the polynomials $p_n(x)$ 
in terms of terminating hypergeometric series; for their notation we 
follow that of standard books~\cite{Andrews,Bailey,Slater}.
The polynomials that will appear here are the Charlier polynomials $C_n(x;a)$, defined by~\cite{Koekoek,Ismail,Andrews}:
\begin{equation}
C_n(x;a) = {\ }_2F_0\left( \myatop{-n,-x}{-} ; -\frac{1}{a} \right). \label{Charlier}
\end{equation}
These polynomials satisfy a discrete orthogonality relation:
\begin{equation}
\sum_{x=0}^{\infty} \frac{a^x}{x!} C_m(x;a) C_n(x;a) = a^{-n} \ee^a n! \delta_{mn}
\label{C-orth}
\end{equation}
when $a>0$.
Using the known forward and backward shift operator relations~\cite{Koekoek}
\begin{align}
& C_n(x;a)-C_n(x-1;a) = -\frac{n}{a} C_{n-1}(x-1;a), \nonumber\\
& C_n(x;a)- \frac{x}{a} C_n(x-1,a) = C_{n+1}(x;a), 
\label{shift}
\end{align}
it is easy to see that the following holds:

\begin{prop}
\label{prop1}
For $\gamma\ne 0$, the solution of the recurrence relations~\eqref{def-px} is given by
\begin{align}
& p_{2n}(x) = \frac{(-\gamma)^{n}}{\sqrt{n!}} C_n(x^2;\gamma^2), \nonumber\\
& p_{2n+1}(x) = - \frac{(-\gamma)^{n-1}}{\sqrt{n!}}\, x\, C_n(x^2-1;\gamma^2). \label{p-solution}
\end{align}
\end{prop}

This result shows, by the way, that the context of Charlier polynomials as overlap coefficients is of a completely different
nature than in~\cite{Vinet2011} (apart from the fact that the overlaps themselves are already different).
In~\cite{Vinet2011}, the recurrence for the overlap coefficients coincides with the common three term recurrence relations for
the Charlier polynomials.
Here, the recurrence relation splits in ``even and odd parts'', and turns out to coincide with shift operator relations~\eqref{shift}
for Charlier polynomials.

Using the orthogonality relation of Charlier polynomials~\eqref{C-orth}, one can deduce the
corresponding orthogonality relation for the polynomials $p_n(x)$.
For Charlier polynomials, the support is discrete and running from 0 to $+\infty$; but due to the
relations~\eqref{p-solution} the support for the $p_n$'s is from $-\infty$ to $+\infty$ (also discrete).

\begin{prop}
\label{prop2}
The polynomials $p_n(x)$ satisfy a discrete orthogonality relation:
\begin{equation}
\sum_{x\in \mathbb{S}} w(x) p_n(x)p_m(x) = \ee^{\gamma^2} \delta_{mn}, \label{ort}
\end{equation}
where
\begin{equation}
\mathbb{S}=\{ \pm \sqrt{k} \ |\  k \in {\mathbb Z}_+ \} = \{\ldots,-\sqrt{3},-\sqrt{2}, -1,0,1,\sqrt{2},\sqrt{3}, \ldots\}, \label{S}
\end{equation}
and where the weight function is given by 
\begin{equation}
w(x) = \frac{1}{2} \frac{\gamma^{2k}}{k!} \qquad \hbox{for } x=\pm\sqrt{k} \quad(k=1,2,3,\ldots)
\label{w}
\end{equation}
and by $w(x)=1$ for $x=0$.
\end{prop} 

So the spectrum of the operator $\hat q$ is discrete and given by~\eqref{S}.

For further use, it will be convenient to introduce the corresponding orthonormal functions defined by 
$\tilde p_n(x)= \ee^{-\gamma^2/2} \sqrt{w(x)} p_n(x)$, i.e.\
\begin{align}
\tilde p_{2n}(x)&= (-1)^n \frac{\sqrt{1+\delta_{x,0}}}{\sqrt{2}}\ee^{-\gamma^2/2} \frac{\gamma^{n+x^2}}{\sqrt{n!(x^2)!}} C_n(x^2;\gamma^2) \nonumber\\
\tilde p_{2n+1}(x)&= \frac{(-1)^n}{\sqrt{2}}\ee^{-\gamma^2/2} \frac{\gamma^{n+x^2-1}}{\sqrt{n!(x^2)!}} x C_n(x^2-1;\gamma^2),
\label{tildep}
\end{align}
satisfying
\begin{equation}
\sum_{x\in\mathbb{S}} \tilde p_m(x) \tilde p_n(x) = \delta_{mn}.
\end{equation}
Following~\eqref{vx}, the normalized eigenvectors of $\hat q$, for an eigenvalue $x\in\mathbb{S}$, are given by
\begin{equation}
\tilde v(x) = \sum_{n=0}^\infty \tilde p_n(x) |n\rangle.
\label{norm-vx}
\end{equation}

\section{A discrete oscillator model related to $\mathfrak{sh}(2|2)$}

In recent years, there has been some interest in the development of new oscillator models,
particularly in the context of finite oscillators.
Let us briefly repeat the general framework here.
For these models, one requires the same dynamics as for the classical or quantum oscillator, 
but the operators corresponding to position, momentum and Hamiltonian can be elements of some algebra different
from the traditional Heisenberg (or oscillator) Lie algebra.
In the one-dimensional case, there are three (essentially self-adjoint) operators involved: 
the position operator $\hat q$, its corresponding momentum operator $\hat p$ and
the Hamiltonian $\hat H$ which is the generator of time evolution. 
The main requirement is that these operators should satisfy the Hamilton-Lie equations 
(or the compatibility of Hamilton's equations with the Heisenberg equations):
\begin{equation}
[\hat H, \hat q] = -\ii \hat p, \qquad [\hat H,\hat p] = \ii \hat q,
\label{Hqp}
\end{equation}
in units with mass and frequency both equal to~1, and $\hbar=1$.
Contrary to the canonical case, the commutator $[\hat q, \hat p]=\ii$ is not required. 
Apart from~\eqref{Hqp} and the self-adjointness, it is then common to require the following conditions~\cite{Atak2001}:
\begin{itemize}
\item all operators $\hat q$, $\hat p$, $\hat H$ belong to some Lie algebra or Lie superalgebra $\cal A$;
\item the spectrum of $\hat H$ in (unitary) representations of $\cal A$ is equidistant.
\end{itemize}
It is within this framework that the oscillator models mentioned in the Introduction fit.
In the current section, we shall examine the oscillator model related to $\mathfrak{sh}(2|2)$ and describe some
of its properties.

\subsection{The $\mathfrak{sh}(2|2)$ oscillator model and its position and momentum wavefunctions}

Clearly, it is possible to construct a new oscillator model using the superalgebra $\mathfrak{sh}(2|2)$ (or,
more precisely, ${\cal S}$) and its representation $V$.
For the Hamiltonian $\hat H$, one can in fact choose a diagonal operator with the same spectrum as the 
canonical oscillator. For this purpose, one extends the superalgebra $\mathfrak{sh}(2|2)$ by a parity or reflection operator $R$,
with action
\[
R|n\rangle = (-1)^n | n\rangle.
\]
Taking then as Hamiltonian
\begin{equation}
\hat H = 2H+\frac12 R
\label{Ham}
\end{equation}
one finds indeed the spectrum
\[
\hat H |n\rangle = (n+\frac12) |n\rangle.
\]
As in~\cite{Jafarov2012}, the position operator $\hat q$ is (up to an overall factor) the most general self-adjoint odd
element of ${\cal S}$. In other words, this is the operator~\eqref{hatq} 
$\hat q = \gamma F^+ + Q^+ + \gamma F^- + Q^-$
investigated in the previous section.
Then the momentum operator must take the form
\begin{equation}
\hat p = \ii \gamma F^+ + \ii Q^+ -\ii \gamma F^- - \ii Q^-,
\label{hatp}
\end{equation}
and~\eqref{Hqp} is satisfied.

In the previous section, we have determined the spectrum $\mathbb{S}$ of the position operator
$\hat q$ and its formal eigenvectors $v(x)$ for $x\in\mathbb{S}$.
The Jacobi matrix corresponding to $\hat p$ is very similar, and one obtains the following result:
\begin{prop}
The spectrum of the momentum operator $\hat p$ is also given by $\mathbb{S}$. For an eigenvalue $y\in\mathbb{S}$,
the normalized eigenvector $\tilde u(y)$ is given by
\begin{equation}
\tilde u (y) = \sum_{n=0}^\infty \ii^n \tilde p_n(y) |n\rangle,
\label{norm-uy}
\end{equation}
where the expressions $\tilde p_n$ are the same as before, given by~\eqref{tildep}.
\end{prop}

We can now continue to investigate some properties of the $\mathfrak{sh}(2|2)$ oscillator model.
First of all, 
the position (resp.\ momentum) wavefunctions of this oscillator are the overlaps 
between the (normalized) $\hat q$-eigenvectors~\eqref{norm-vx} (resp.\ $\hat p$-eigenvectors~\eqref{norm-uy})
and the $\hat H$-eigenvectors. 
Clearly, it is sufficient to consider only the position wavefunctions, denoted by $\varphi_n^{(\gamma)}(x)$
and given by
\begin{equation}
\varphi_n^{(\gamma)}(x) = \tilde p_n(x), \label{varphi}
\end{equation}
where $x$ belongs to $\mathbb{S}$.
The dependence of the wavefunctions on the model parameter $\gamma$ is made apparent in the notation.
So we are dealing with a discrete position wavefunction (but with infinite support).
Wavefunctions of this type have appeared also for the (infinite discrete) $\mathfrak{sl}(2|1)$ oscillator model~\cite{Jafarov2012}.
In the current case, the shape of these discrete position wavefunctions is similar.
We have plotted some examples of $\varphi_n^{(\gamma)}(x)$ in Figure~1, for certain values of $\gamma$.
Note that the shapes look indeed like discrete versions of the common oscillator wavefunctions in terms
of normalized Hermite polynomials.

As far as the plots in Figure~1 are concerned, keep in mind that we are dealing with discrete wavefunctions
with infinite support $\mathbb{S}$. So only the dots in Figure~1 are part of the wavefunctions, not the curves connecting
these dots (these curves have been drawn just to emphasize the ``shape'' of the discrete wavefunctions).
Note also that we have plotted only a finite part of the infinite support (namely only the dots corresponding
to the support $[-\sqrt{k},+\sqrt{k}]$ with $k=10$). 

\subsection{A limit related to the position wavefunction}

It is not straightforward to consider direct limits of the $\mathfrak{sh}(2|2)$ position wavefunctions $\varphi^{(\gamma)}_n(x)$,
e.g.\ when the model parameter $\gamma$ varies.
Instead, we need to look for another oscillator model of which the current model itself is a limit.
This is provided by the finite oscillator model related to the Lie superalgebra $\mathfrak{sl}(2|1)$~\cite{JV2012}.
Recall that this $\mathfrak{sl}(2|1)$ oscillator model has two parameters: a representation parameter $j$ (which is
a nonnegative integer) and a model parameter $p$ (with $0<p<1$). The representations in~\cite{JV2012} are
finite-dimensional (with dimension $2j+1$). The support of the $\mathfrak{sl}(2|1)$ position wavefunctions is given by $\pm\sqrt{k}$ 
(where $k=0,1,\ldots,j$). The actual $\mathfrak{sl}(2|1)$ wavefunctions are given by~\cite[(44)-(48)]{JV2012}:
\begin{align}
& \phi^{(p,j)}_{2n} (x) = (-1)^n \sqrt{\frac{1+\delta_{x,0}}{2}} \tilde K_n(x^2;p,j),\qquad (n=0,1,\ldots,j), \nonumber\\
& \qquad x=\pm\sqrt{k},\qquad k=0,1,\ldots j; \nonumber\\
& \phi^{(p,j)}_{2n+1} (x) = \pm (-1)^n \sqrt{\frac{1-\delta_{x,0}}{2}} \tilde K_n(x^2-1;p,j-1),\qquad (n=0,1,\ldots,j-1), \nonumber\\
& \qquad x=\pm\sqrt{k},\qquad k=0,1,\ldots j. \nonumber
\end{align}
Herein, $\tilde K_n(x;p,N)$ is the normalized Krawtchouk polynomial~\cite[(24)]{JV2012}, i.e.\ the Krawtchouk polynomial
$K_n(x;p,N)$ \cite[9.11]{Koekoek} multiplied by the square root of its weight function and the inverse of its norm.

In general, the representation parameter $j$ and the parameter $p$ are independent in the $\mathfrak{sl}(2|1)$ oscillator model.
Let us, however, consider the following dependence:
\begin{equation}
p=\gamma^2/j,
\label{pj}
\end{equation}
where $\gamma$ is a nonzero constant. Then the classical limit from Krawtchouk polynomials to Charlier polynomials
\cite[9.14]{Koekoek} implies:
\begin{equation}
\lim_{j\rightarrow \infty} \phi^{(\gamma^2/j,j)}_{n} (x) = \varphi^{(\gamma)}_n(x).
\label{limit}
\end{equation}
In other words, for large values of $j$ the position wavefunctions of the $\mathfrak{sl}(2|1)$ finite oscillator model tends to 
those of the $\mathfrak{sh}(2|2)$ oscillator model,
provided the parameter $p$ is represented as $\gamma^2/j$ in the limit.

The behaviour of these wavefunctions for large values of $j$ is also confirmed in plots.
In Figure~2, we have plotted the ground state ($n=0$) and first excited state ($n=1$) of the $\mathfrak{sl}(2|1)$ oscillator
for increasing values of $j$, namely $j=5, 10, 20, 30$ (note that from $j=20$ onwards, not the complete finite wavefunction is
plotted, since we plot the support only in the interval $[-4,4]$). 
For these plots, we have taken $\gamma=1$ as an example.
In the bottom line of Figure~2, we have plotted the corresponding wavefunctions of the $\mathfrak{sh}(2|2)$ oscillator,
again for the value $\gamma=1$ (and again in the interval $[-4,4]$). 
From these plots, the limit relation is clearly illustrated.

Recall from~\cite{JV2012} that for large $j$-values the behaviour of the $\mathfrak{sl}(2|1)$ wavefunctions
$\phi^{(p,j)}_{n} (x)$ is the same as the behaviour of paraboson wavefunctions $\Psi^{(2p\alpha-1)}_n(x)$ for large values
of $\alpha$. In the process just described, however, $p$ is a fixed constant.
This explains why it is not possible to give a direct limit from the $\mathfrak{sl}(2|1)$ wavefunctions to a known 
family of continuous wavefunctions like the paraboson wavefunctions or the canonical oscillator wavefunctions.

\subsection{An expression of the $\mathfrak{sh}(2|2)$ Fourier transform}

In canonical quantum mechanics, the momentum wavefunction (in $L^2({\mathbb R})$) is given by the Fourier transform of
the position wavefunction (and vice versa), with kernel $K(x,y)$:
\[
\psi(y)=  \int K(x,y) \phi(x)dx, \qquad K(x,y)=\frac{1}{\sqrt{2\pi}} \ee^{-\ii xy}.
\]
This means that $K(x,y)$ is also the overlap of the (formal) position eigenvector for the eigenvalue $x$ with the
momentum eigenvector for the eigenvalue $y$. 
In the current case, the kernel for the corresponding $\mathfrak{sh}(2|2)$ Fourier transform is
\begin{equation}
K^{(\gamma)}(x,y) = \langle \tilde v(x), \tilde u(y)\rangle   
= \sum_{n=0}^\infty (-\ii)^n \varphi^{(\gamma)}_n(x) \varphi^{(\gamma)}_n(y),
\label{K}
\end{equation}
where $\varphi^{(\gamma)}_n(x)=\tilde p_n(x)$ is given by~\eqref{tildep}.

It is easy to compute this kernel explicitly. Splitting the sum over even and odd $n$-values, using~\eqref{tildep},
and using the known bilinear generating function for Charlier polynomials \cite{Meixner39},
\begin{equation}
\sum_{m=0}^\infty \frac{t^m}{m!} C_m(x;a) C_m(y;a) = \ee^t (1-t/a)^{x+y} 
{\ }_2F_0\left( \myatop{-x,-y}{-} ; \frac{t}{(a-t)^2} \right),
\end{equation}
one finds
\begin{align}
K^{(\gamma)}(x,y) & =\frac{\ee^{-2\gamma^2} (2\gamma)^{x^2+y^2}}{2\sqrt{(x^2)!(y^2)!}} 
\left[ \sqrt{(1+\delta_{x,0})(1+\delta_{y,0})} {\ }_2F_0\left( \myatop{-x^2,-y^2}{-} ; -\frac{1}{4\gamma^2} \right) \right.\nonumber\\
& \left. -\ii \frac{xy}{4\gamma^2} {\ }_2F_0\left( \myatop{1-x^2,1-y^2}{-} ; -\frac{1}{4\gamma^2} \right)\right].
\end{align}
Herein, $x$ and $y$ are elements of $\mathbb{S}$, i.e.\ they are of the form $x=\pm\sqrt{k}, y=\pm\sqrt{l}$ with
$k$ and $l$ nonnegative integers.

\subsection{Some other physical quantities of the $\mathfrak{sh}(2|2)$ oscillator}

As we are speaking of an oscillator model, let us compute here some other quantities that play
a role in physics (all under the assumption $m=\omega=\hbar=1$).
Taking into account the expression for the position operator $\hat q$ \eqref{hatq} and the momentum operator $\hat p$
\eqref{hatp}, one finds
\[
[\hat q, \hat p ] = -2\ii (\gamma^2 [F^+,F^-]+ [Q^+,Q^-]).
\]
Using the actions~\eqref{act1} in the Fock space, this leads to ($n=0,1,2,\ldots$)
\[
\left[ {\hat q,\hat p} \right]\left| n \right\rangle = 2\ii \left[ {\left( { - 1} \right)^n \left( {\gamma ^2 - \frac{1}{2}n} \right) + \frac{{1 - \left( { - 1} \right)^n }}{4}} \right]\left| n \right\rangle ,
\]
or for even and odd states separately:
\begin{align}
& [\hat q, \hat p ] |2k\rangle = 2\ii (\gamma^2-k) |2k\rangle, \nonumber\\
& [\hat q, \hat p ] |2k-1\rangle = -2\ii (\gamma^2-k) |2k-1\rangle.
\end{align}
These relations are similar to the ones for the paraboson oscillator~\cite[(16)]{JLV2008}, apart from the $k$-dependence.

Another quantity that can easily be computed is the standard deviation for $\hat q$ and $\hat p$ when
the system is in the stationary state $|n\rangle$. 
Using
\[
(\Delta\hat q)_n = \sqrt{ \langle n| \hat q^2 | n\rangle - \langle n| \hat q |n \rangle^2}
\]
and the explicit action of $\hat q = \gamma F^+ + Q^+ + \gamma F^- + Q^-$, one finds
\[
(\Delta\hat q)_n = \sqrt{\gamma^2 +\frac{1}{2} n +\frac{1}{4}(1-(-1)^n)}.
\]
The computation for $(\Delta\hat p)_n$ is the same, so one finds as ``uncertainty relation'' quantity
\[
(\Delta\hat q)_n(\Delta\hat p)_n = \gamma^2 +\frac{1}{2} n +\frac{1}{4}(1-(-1)^n).
\]
Although this depends on the state number, note that its value is the same for two consecutive states:
\[
\left( {\Delta \hat q} \right)_{2k - 1} \left( {\Delta \hat p} \right)_{2k - 1} 
= \left( {\Delta \hat q} \right)_{2k} \left( {\Delta \hat p} \right)_{2k} =  \gamma ^2 + k.
\]

Also the operator $\frac{{\hat p}^2}{2}+\frac{{\hat q}^2}{2}$ is worth considering.
On the algebraic level, one finds
\[
\frac{{\hat p}^2}{2}+\frac{{\hat q}^2}{2} = \gamma^2+H,
\]
following from the explicit expressions of $\hat p$, $\hat q$ and~\eqref{oddodd}.
For the action in the Fock space, this implies
\[
(\frac{{\hat p}^2}{2}+\frac{{\hat q}^2}{2}) |n\rangle = 
(\gamma^2+\frac{n}{2}+\frac{1-(-1)^n}{4}) |n\rangle.
\]

Observe that all these quantities yield minor deviations from the corresponding ones for the canonical
oscillator or for the paraboson oscillator.

\section{Some concluding remarks}

Before coming to some conclusions, let us mention a differential operator realization of the operators studied here.
It is well known that the Heisenberg-Weyl superalgebra and its Fock representation can be realized in
terms of an ordinary variable $z$ and a Grassmann variable $\theta$ (with $\theta^2=0$). The Fock representation states
are then given by
\begin{equation}
|2n\rangle = \frac{z^n}{\sqrt{n!}}, \qquad |2n+1\rangle = \frac{z^n}{\sqrt{n!}} \theta.
\end{equation}
The Lie superalgebra basis elements take the form
\begin{align}
& F^+=\theta,\ F^-=\partial_\theta,\ Q^+ = z\partial_\theta,\ Q^- = \theta\partial_z, \nonumber\\
& E^+=z,\ E^-=\partial_z,\ H=z\partial_z +\theta\partial_\theta.
\end{align}
In this realization, the position operator corresponds to the differential operator
\begin{equation}
\hat q = \theta (\gamma +\partial_z) +\partial_\theta (\gamma+z).
\label{hatq-real}
\end{equation}
It is worthwhile to give the eigenfunctions of this operator explicitly.
This can be done using the normalized eigenfunctions~\eqref{norm-vx} of $\hat q$ and by using the generating function for 
Charlier polynomials~\cite{Koekoek}
\[
\sum_{n=0}^\infty C_n(x;a) \frac{t^n}{n!} = \ee^t(1-\frac{t}{a})^x.
\]
One finds:
\begin{align}
\tilde v(x) & = \frac{1}{\sqrt{2}} \frac{\ee^{-\gamma^2/2-\gamma z}}{\sqrt{k!}} (\gamma+z)^{k-1} 
(\gamma+z \pm \sqrt{k} \theta) \nonumber\\
& \hbox{ for } x=\pm\sqrt{k}, \qquad k=1,2,3,\ldots
\end{align}
and $\tilde v(0)=\ee^{-\gamma^2/2-\gamma z}$. 
In this realization, it is easy to verify that $\hat q \tilde v(x) = x \tilde v(x)$, for $x\in\mathbb{S}$.

To summarize, we have worked out a new model for a quantum oscillator based on the Lie superalgebra $\mathfrak{sh}(2|2)$
and its Fock representation. 
This is the second algebraic oscillator model (in the sense of the introduction of section~4) based
on a Lie superalgebra (the other one being based on $\mathfrak{sl}(2|1)$).
The Hamiltonian $\hat H$, the position $\hat q$ and the momentum $\hat p$ of the model are three self-adjoint
elements of the enveloping algebra of $\mathfrak{sh}(2|2)$ satisfying the Hamilton-Lie equations~\eqref{Hqp}.
The model allows a free parameter $\gamma$ which appears in the expression of $\hat q$.
The spectrum of the energy operator $\hat H$ coincides with that of the canonical quantum oscillator.
Apart from the presentation of the new $\mathfrak{sh}(2|2)$ oscillator model, our main
result is the determination of the spectrum of the position operator and its wavefunctions.
The spectrum of the position operator $\hat q$ is infinite discrete and given by
$\mathbb{S}= \{\ldots,-\sqrt{3},-\sqrt{2}, -1,0,1,\sqrt{2},\sqrt{3}, \ldots\}$.
The most interesting fact is that the position wavefunctions $\varphi^{(\gamma)}_n(x)$ correspond
to ``symmetrized'' Charlier polynomials (i.e.\ even wavefunctions are Charlier polynomials in $x^2$, and
odd wavefunctions are $x$ times Charlier polynomials in $x^2$).
Some plots of the position wavefunctions $\varphi^{(\gamma)}_n(x)$ reveal interesting properties.
In particular, the wavefunctions can be seen as limits of finite discrete wavefunctions appearing
in an $\mathfrak{sl}(2|1)$ oscillator model.

The model presented here is elegant and simple from a mathematical point of view. 
Its relevance does not come directly from a problem in physics. 
On the other hand, harmonic oscillators are fundamental in many branches of physics, 
so we believe that all oscillator models wich are minor deviations from the canonical oscillator deserve 
attention.
The current model is sufficiently simple, and could perhaps be extended to more dimensions, e.g.\ using
the supersymmetric Heisenberg-Weyl algebra $\mathfrak{sh}(2|d)$. 
Whether such an extension gives rise to interesting results related to orthogonal polynomials,
is a question that will be studied in the future.

\section*{Acknowledgments}
E.I.~Jafarov kindly acknowledges support from TWAS Research Grant 11-085, Research Grant EIF-2012-2(6)-39/08/1 of the Science Development Foundation under the President of the Republic of Azerbaijan and support from ICTP within the regular associateship scheme.

\newpage
\begin{figure}[th]
\[
\begin{tabular}{cccc}
\hline\\[-3mm]
 & $\gamma=0.5$ & $\gamma=1$ & $\gamma=2$ \\[1mm]
$n=0$&\includegraphics[scale=0.50]{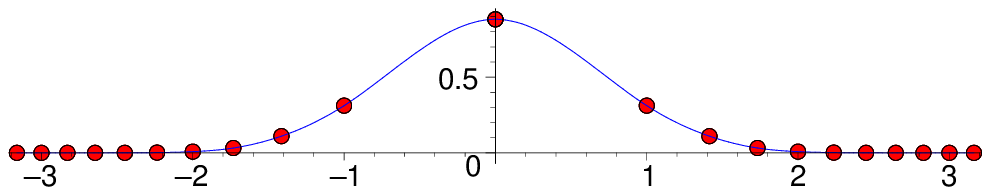} & \includegraphics[scale=0.50]{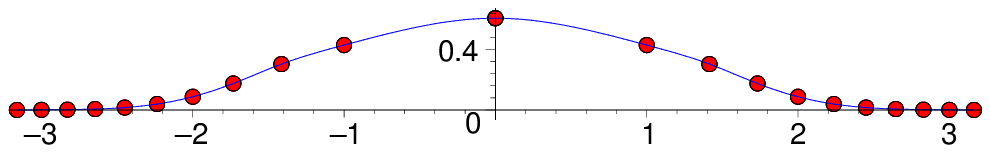} & \includegraphics[scale=0.50]{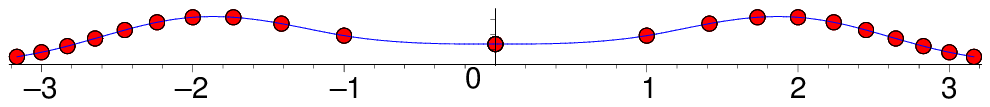} \\[10mm]
$n=1$&\includegraphics[scale=0.50]{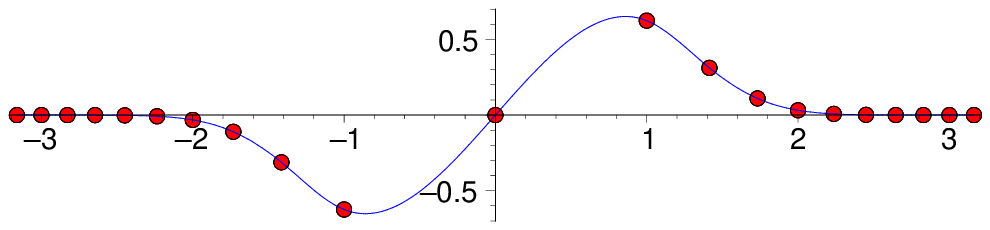} & \includegraphics[scale=0.50]{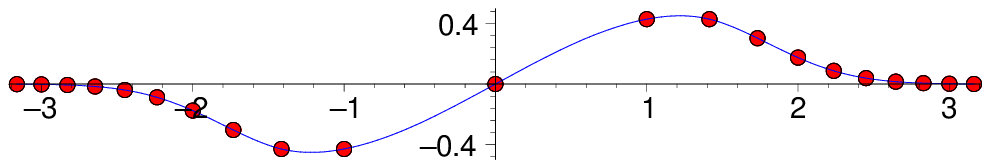} & \includegraphics[scale=0.50]{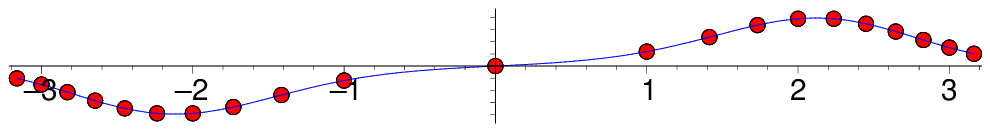} \\[10mm]
$n=2$&\includegraphics[scale=0.50]{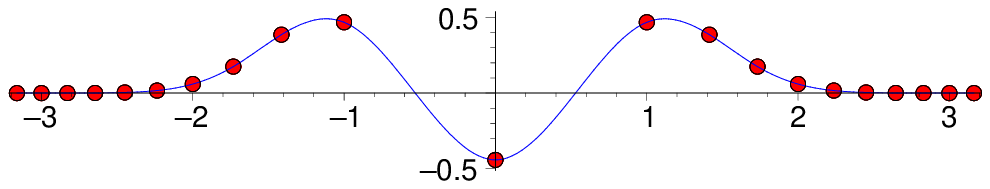} & \includegraphics[scale=0.50]{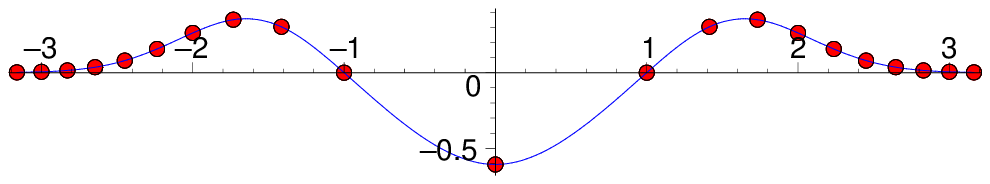} & \includegraphics[scale=0.50]{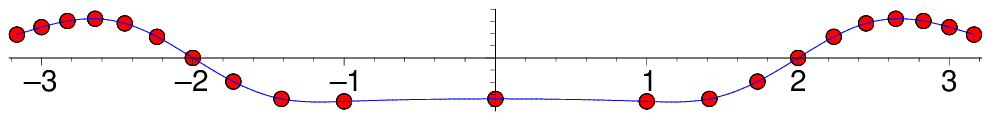} \\[10mm]
$n=3$&\includegraphics[scale=0.50]{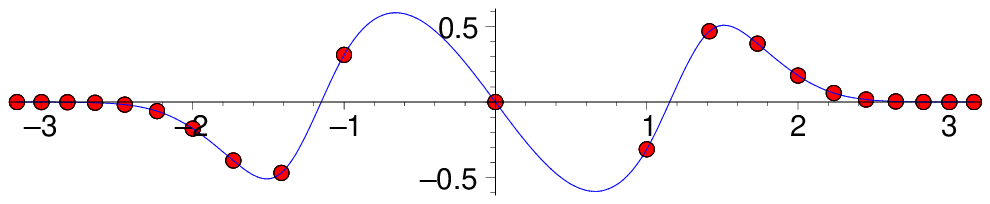} & \includegraphics[scale=0.50]{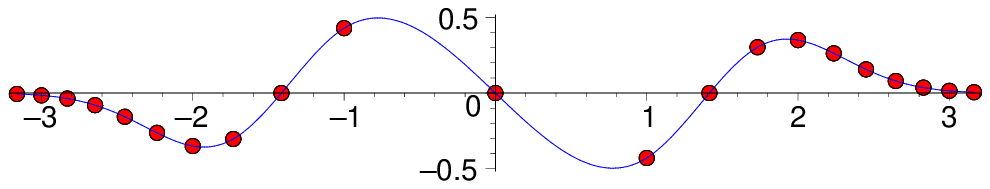} & \includegraphics[scale=0.50]{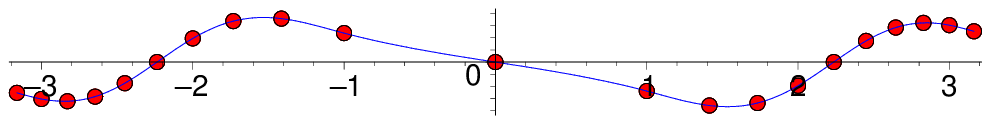} \\
\hline
\end{tabular} 
\]
\caption{Plots of (part of) the wavefunctions $\varphi^{(\gamma)}_n(x)$ for $n=0,1,2,3$ 
and for some values of $\gamma$, where $x\in\mathbb{S}$.}
\label{fig1}
\end{figure}

\newpage
\begin{figure}[th]
\[
\begin{tabular}{ccc}
\hline\\[-3mm]
 & $n=0$ & $n=1$  \\[1mm]
$j=5$&\includegraphics[scale=0.70]{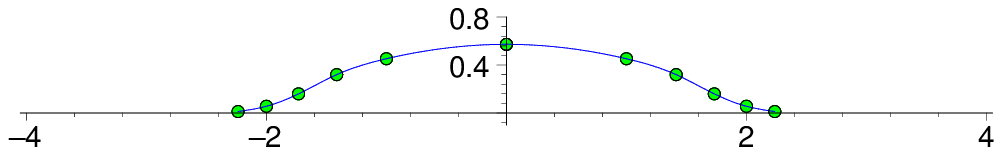} & \includegraphics[scale=0.70]{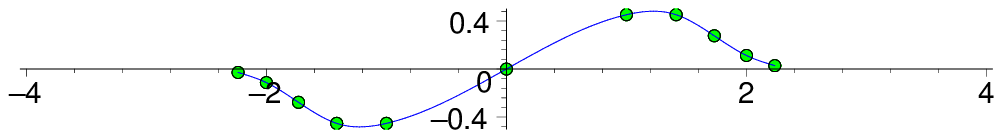} \\[10mm]
$j=10$&\includegraphics[scale=0.70]{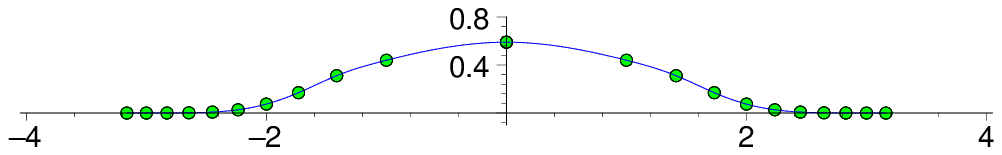} & \includegraphics[scale=0.70]{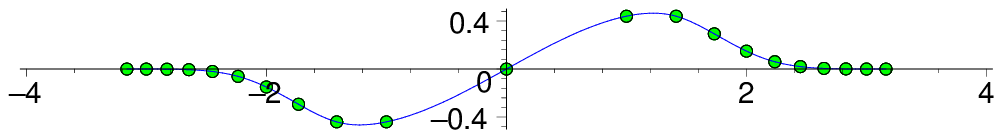} \\[10mm]
$j=20$&\includegraphics[scale=0.70]{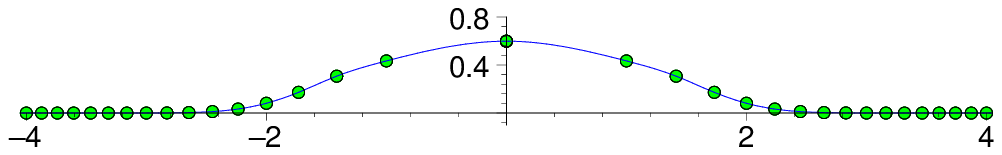} & \includegraphics[scale=0.70]{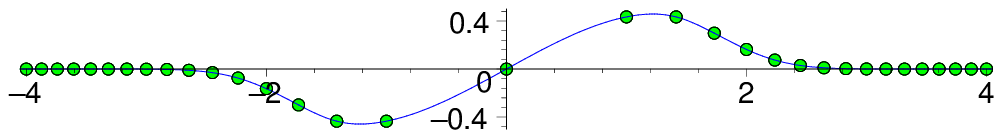} \\[10mm]
$j=30$&\includegraphics[scale=0.70]{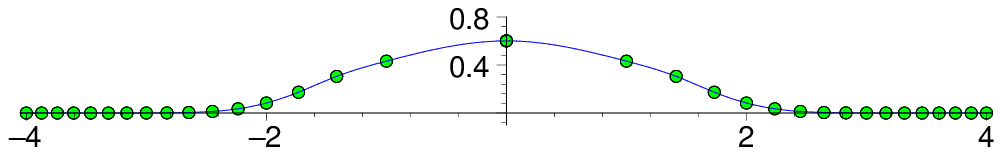} & \includegraphics[scale=0.70]{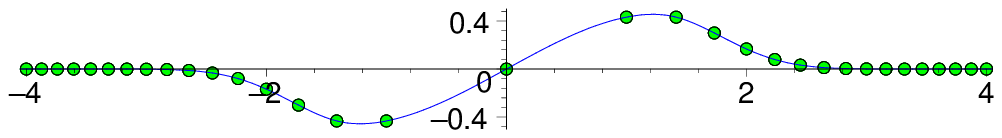} \\[15mm]
$\varphi^{(\gamma)}_n(x)$&\includegraphics[scale=0.70]{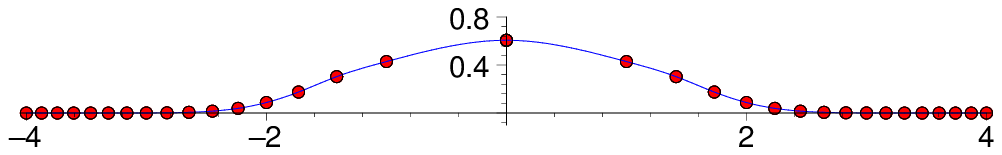} & \includegraphics[scale=0.70]{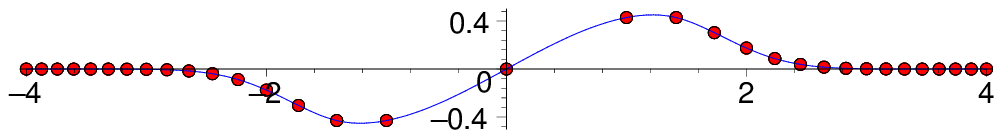}  \\
\hline
\end{tabular} 
\]
\caption{Plots of the $\mathfrak{sl}(2|1)$ wavefunctions $\phi^{(p,j)}_n(x)$ for $n=0,1$ with $p=\gamma^2/j$
and for some values of $j$ (with $\gamma=1$), and their comparison with $\varphi^{(\gamma)}_n(x)$ (bottom figure).}
\label{fig2}
\end{figure}

\end{document}